\documentclass[lettersize,journal]{IEEEtran}
\usepackage{amsmath,amsfonts}
\usepackage{algorithmic}
\usepackage{algorithm}
\usepackage{array}
\usepackage[caption=false,font=normalsize,labelfont=sf,textfont=sf]{subfig}
\usepackage{textcomp}
\usepackage{stfloats}
\usepackage{url}
\usepackage{verbatim}
\usepackage{graphicx}
\usepackage[nocompress]{cite}
\usepackage[table]{xcolor}
\usepackage{booktabs}
\usepackage{makecell}
\usepackage{pifont}
\usepackage{orcidlink} 
\usepackage{textcomp} 

\hypersetup{colorlinks=true, linkcolor=blue, citecolor=blue, urlcolor=blue}

\begin{document}

\title{Can We Hear from Events? Generating Speech from Event Camera}

\author{Jingping~Fang$^{1*}$, 
        Lin~Chen$^{1*}$, 
        Chenyang~Xu$^{2}$, 
        Tong~Zhao$^{3}$, 
        Weidong~Cai$^{4}$, 
        and~Xiaoming~Chen$^{1\dagger}$% 

\thanks{*Equal Contribution; $^{\dagger}$Corresponding Author.}%
\thanks{$^{1}$Beijing Technology and Business University.}%
\thanks{$^{2}$Xidian University.}%
\thanks{$^{3}$Tongji University.}%
\thanks{$^{4}$University of Sydney.}%
}

% The paper headers
%\markboth{IEEE Transactions on Multimedia,~Vol.~14, No.~8, May~2026}%
%{Shell \MakeLowercase{\textit{et al.}}: A Sample Article Using IEEEtran.cls for IEEE Journals}

%\IEEEpubid{0000--0000/00\$00.00~\copyright~2021 IEEE}
% Remember, if you use this you must call \IEEEpubidadjcol in the second
% column for its text to clear the IEEEpubid mark.
\maketitle
\begin{abstract}
Traditional RGB-based speech generation faces Temporal Granularity Mismatch since fixed camera exposure times inevitably blur the high-frequency articulatory transients essential for rendering emotional speech. To break this ceiling, we propose EventSpeech as a novel text-conditioned framework pioneering the use of neuromorphic events for expressive speech generation, since these microsecond-precise events naturally align with acoustic waveform dynamics. Our architecture integrates a dedicated Event Encoder to model sparse neuromorphic events alongside a multi-scale Audio Encoder featuring a Hierarchical Wavelet Contextualizer (HWC). A bidirectional alignment mechanism seamlessly synchronizes linguistic content and visual dynamics with dense acoustic features. Furthermore, we construct EVT-SPK as the first benchmark comprising large-scale synthetic data and real-world recordings from specialized neuromorphic hardware. Extensive evaluations demonstrate that EventSpeech significantly outperforms current baselines by preserving fine-grained emotions and resisting motion blur to establish a new paradigm for multimodal speech generation. Code and demo are available at \href{https://xrfang-0102.github.io/EventSpeechWeb/}{https://xrfang-0102.github.io/EventSpeechWeb/}.

\end{abstract}


\section{Introduction}
Achieving human-level expressiveness in speech generation remains a formidable challenge in machine learning. While Traditional text-to-speech (TTS) \cite{mehta2024matcha,tan2024naturalspeech,kim2021conditional}    and vision-assisted \cite{lee2023imaginary,wang2024v2a} models have made strides in intelligibility, they often produce speech that is over-smoothing, lacking the sharp prosodic nuances and emotional transients of real human speech. Existing research typically attributes this limitation to model capacity\cite{chen2024vall}, text-audio alignment\cite{kim2020glow}, or training strategies\cite{mehta2024matcha}. However, human speech is fundamentally a multimodal physical process, where acoustic expressiveness is tightly coupled with high-frequency facial micro-dynamics\cite{chu2025dcptalk}. While recent audio-visual approaches\cite{cong2024styledubber,lee2023imaginary} attempt to capture this physical grounding, we argue that the bottleneck is fundamentally physical \textit{the reliance on frame-based sensors}.

The Standard RGB video \cite{xing2024seeing,liu2024tell} operates by integrating light intensity over a fixed exposure window (typically 33ms at 30fps \cite{xiao2026learning}). This integration acts as a low-pass filter, irreversibly smoothing out the high-frequency visual jitters (e.g., subtle lip tremors, rapid jaw accelerations, and microexpressions physically coupled with the generation of high frequency acoustic features). We term this phenomenon the \textit{Temporal Granularity Mismatch}. When a model is conditioned on blurred visual inputs from standard frames, it inevitably predicts averaged acoustic features, resulting in speech that lacks temporal crispness and emotional vitality. To generate truly expressive speech, we necessitates shifting the visual modality from time-averaged frame accumulation to asynchronous kinematic sensing.

\begin{figure}[t]
    \centering
    \includegraphics[width=\linewidth, keepaspectratio]{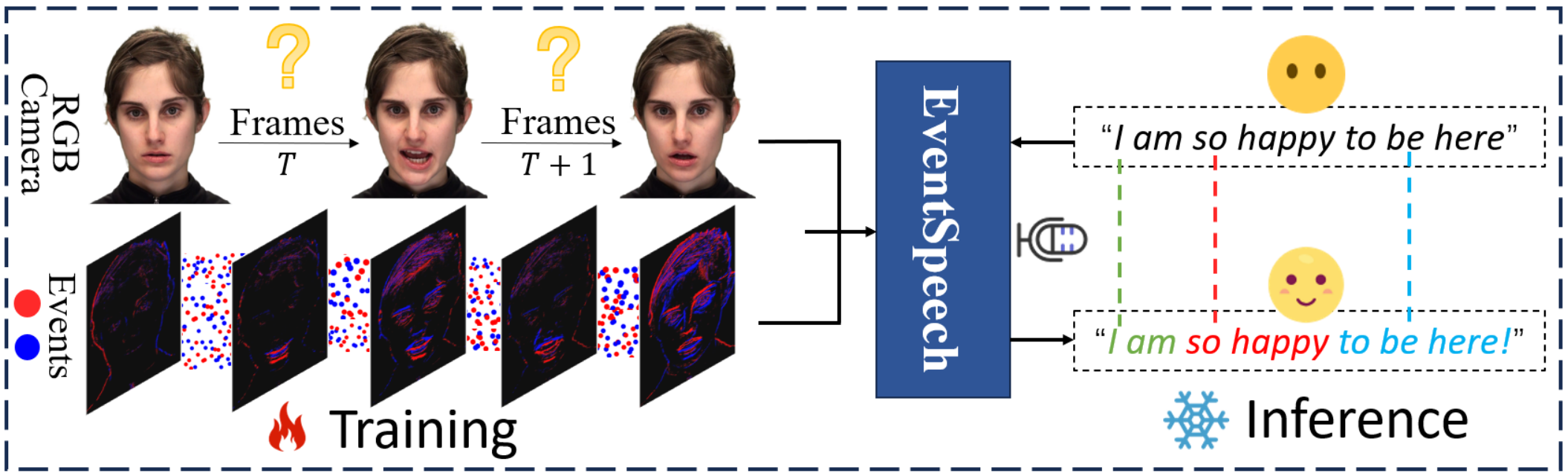}
    \caption{\textbf{Pipeline of EventSpeech.} During training (\textbf{left}), the framework learns highly natural speech representations guided by audio supervision and neuromorphic events, which capture high-frequency articulatory dynamics often missed by standard RGB frames. During inference (\textbf{right}), strict multimodal consistency accommodates both text-only generation via learned alignment priors and vision-augmented synthesis when events are available.}
    \label{fig1}
\end{figure}

In this paper, we propose \textbf{EventSpeech}, a novel framework (Figure \ref{fig1}) that bridges this gap by introducing neuromorphic events to speech generation. Unlike standard RGB cameras, bioinspired event sensors \cite{hu2021v2e,chakravarthi2024recent} asynchronously record intensity changes with microsecond resolution to eliminate motion blur. This mechanism precisely captures facial articulatory kinematics\cite{zhang2025eventlip} to provide continuous motion guidance naturally aligned with acoustic waveforms. To effectively map sparse, asynchronous events to dense, continuous speech, we design a specialized architecture featuring the HWC from Audio Encoder for multi-scale time-frequency modeling, and a bidirectional alignment module that ensures precise synchronization between visual kinematics and acoustic boundaries. By forcing the text encoder to learn a surrogate representation of visual dynamics during training, \textbf{EventSpeech} enables text-driven inference with learned alignment priors. Ultimately, this robust physical foundation enables the model to learn and generate highly expressive speech from continuous visual information, faithfully reproducing the physical rhythms and vivid emotional shifts that prior works have failed to address.

\IEEEpubidadjcol
Furthermore, a critical barrier to event-based speech research has been the reliance on simulated data, while simulated events can resolve specific scenarios\cite{jiang2023event} where standard RGB inputs fail, they inherently lack authentic real world sensor noise\cite{ding2023mlb} and physical illumination dynamics. To address this, we construct \textbf{EVT-SPK}, the first comprehensive benchmark for event-driven speech generation. It encompasses a large scale synthetic corpus derived from high quality emotion datasets\cite{wang2020mead,livingstone2018ryerson} and a real world dataset comprising 2.8k recordings captured using synchronized neuromorphic hardware (\textit{DAVIS346}) and an high fidelity audio recorder (\textit{H3-VR}). These physical recordings deliberately cover demanding environmental conditions (e.g., extreme low light environments and rapid articulatory motions causing severe RGB motion blur). Extensive experiments demonstrate that \textbf{EventSpeech} significantly outperforms state-of-the-art (SOTA), particularly amid rapid articulation and subtle facial dynamics thereby establishing neuromorphic events as a superior modality for authentic speech generation. 

Our main contributions are summarized as follows: 
\begin{enumerate}
    \item We propose \textbf{EventSpeech}, the first framework leveraging neuromorphic events for speech generation. We resolve the Temporal Granularity Mismatch of RGB sensors, achieving microsecond-level alignment between visual kinematics and acoustic waveforms.

    \item We establish a physically-grounded human perception paradigm. Unlike traditional methods limited by RGB camera frames, our event-driven approach captures subtle motion changes more accurately and fully.

    \item We construct \textbf{EVT-SPK}, the first benchmark featuring real-world recordings and semi-automated preprocessing pipeline. Uniquely containing data captured via event sensors, it bridges the simulation gap and validates our model's ability on real-world events.

    \item We achieve outstanding performance against comparative baselines. Objective and subjective results experimentally demonstrate our method's superior robustness against motion blur and fidelity in fine-grained prosody preservation, proving the necessity of neuromorphic events for expressive speech generation.
\end{enumerate}

\section{Related Work}
\label{rework}

\noindent \textbf{Text-Driven Speech Generation.}
Traditional TTS primarily relies on text to synthesize natural speech. The field has evolved from autoregressive sequence-to-sequence architectures like Tacotron\cite{wang2017tacotron} and Tacotron2\cite{shen2018natural} to faster non-autoregressive models such as FastSpeech and FastSpeech2. More recently, generative paradigms have achieved superior fidelity, including diffusion-based models\cite{popov2021grad,kong2020diffwave}, flow-based approaches \cite{kim2020glow,mehta2024matcha},  and fully end-to-end frameworks\cite{kim2021conditional,tan2024naturalspeech}. Recent advances in flow matching\cite{chen2025f5} and diffusion transformers\cite{lee2025ditto} further push the boundaries of quality. However, relying solely on text training inherently neglects non-verbal cues such as facial expressions and lip kinematics, limiting the ability to capture speaker-specific prosodic nuances and necessitating the integration of visual modalities.

\noindent \textbf{Frames-Based Visual Speech Generation.} Integrating visual signals has proven effective for enhancing audio-visual synchronization. Works like VDTTS\cite{hassid2022more} and Seeing What You're Saying\cite{vougioukas2020realistic} utilize RGB frames to modulate speech, while some approaches \cite{lee2023imaginary, wang2024v2a} explicitly extract facial or emotional features to guide prosody. Furthermore, in the specialized domain of visual voice cloning (V2C) and automated movie dubbing approaches\cite{sung2025voicecraft,cong2023learning,cong2024styledubber} have pushed the boundaries of visually-conditioned speech by leveraging hierarchical facial emotions and multi-scale phoneme-level style learning to synthesize highly contextualized, lip-synced prosody. Extending to broader multimodal tasks, recent video-to-audio generation models \cite{zhang2026foleycrafter,luo2023diff,ton2025taro} similarly leverage visual motion cues to enforce fine-grained temporal alignment in general sound synthesis. Despite these advances, we argue that frames-based approaches hit a physical ceiling. Standard cameras integrate light intensity over a fixed exposure window (typically 33ms\cite{vougioukas2020realistic}), causing irreversible motion blur that smooths out high-frequency articulatory transients. Even recent attempts at feature interpolation\cite{hong2025audio,ki2025float} cannot recover these lost micro-dynamics, often resulting in generated speech with over-smoothed prosody that lacks the temporal crispness. 

\noindent \textbf{Neuromorphic Events-Based Audio Generation.} Neuromorphic event cameras fundamentally differ from conventional frame-based sensors by asynchronously recording per-pixel brightness changes with microsecond resolution and an extensive dynamic range\cite{chakravarthi2024recent}. While recent methodologies have successfully utilized these event streams to generate environmental audio including impact sounds from object interactions\cite{rebecq2019high,su2023physics,liu2024tell} and cross-modal alignment tasks\cite{luo2023diff,xing2024seeing}, a critical research gap persists in human speech generation. Unlike environmental noises triggered by discrete physical collisions, speech production involves continuous and non-linear biological kinematics such as subtle lip deformations and micro-tremors that convey dense semantic information. Although event cameras have been integrated into speech-related domains including visual speech recognition\cite{kanamaru2023isolated,tan2022multi,chen2024collaborative} and voice activity detection\cite{arriandiaga2021audio}, these conventional approaches restrict event cues exclusively to discriminative analysis. They inherently fail to address the complex generative challenge of synthesizing expressive speech waveforms directly from neuromorphic signals. Prior to the proposed EventSpeech framework, no existing methodology has successfully mapped sparse and asynchronous events in synergy with textual linguistic content into the continuous latent space of expressive speech. The proposed method bridges this fundamental gap by treating event representations as high-fidelity physical records of articulatory dynamics rather than generic motion triggers to establish a novel paradigm for visually guided TTS generation. 

\section{Framework: EventSpeech}
\label{methods}

\begin{figure*}[htbp]
    \centering
    \includegraphics[width=\linewidth]{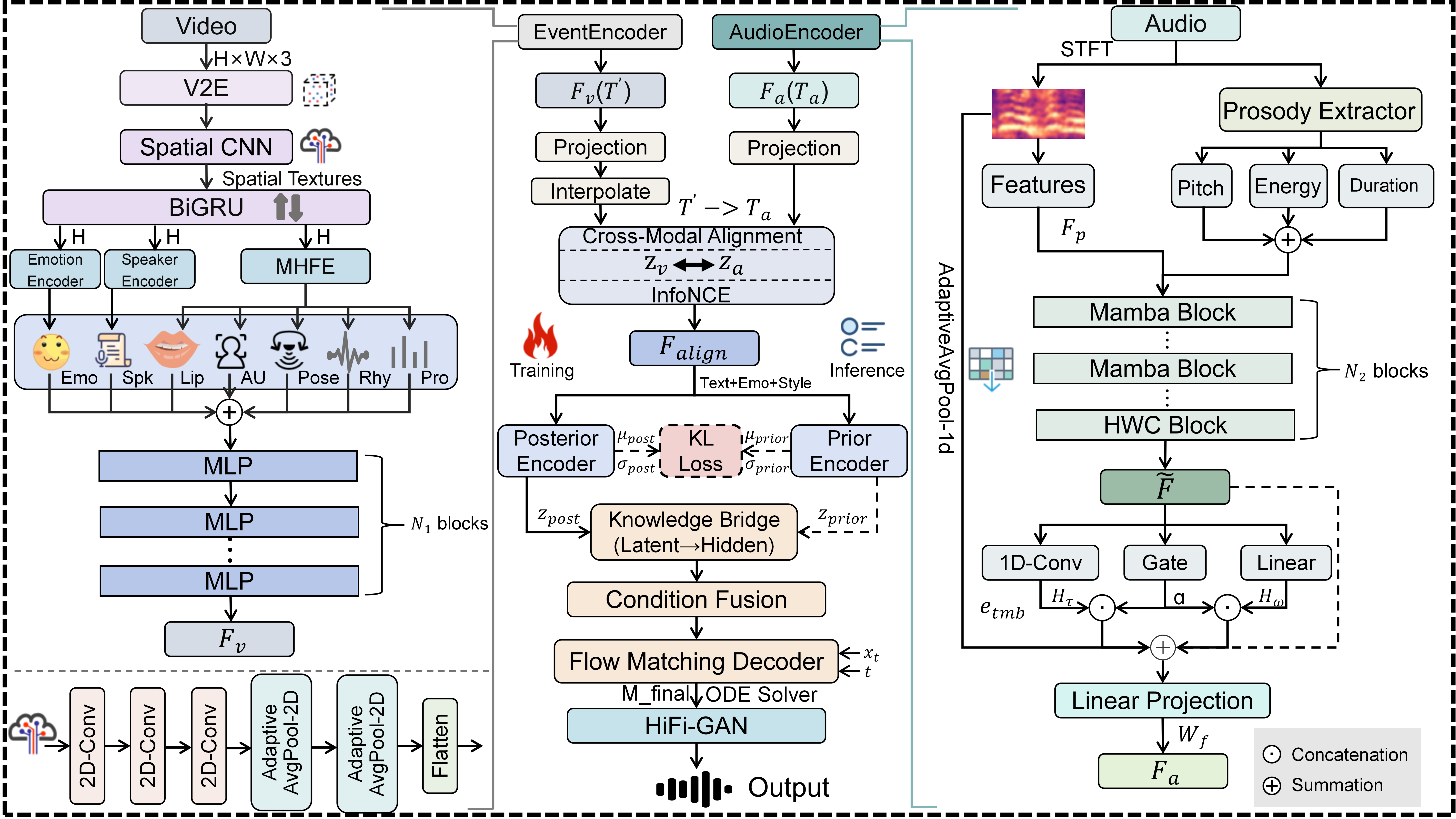}
    \caption{\textbf{EventSpeech Core Architecture.} \textbf{Left:} Event Encoder extracts disentangled kinematics from V2E events via MHFE. \textbf{Right:} Audio Encoder fuses Mamba and HWC blocks for hierarchical acoustic modeling. \textbf{Middle:} Following InfoNCE cross-modal alignment, a VITS dual-stream framework routes latent representations via a Knowledge Bridge to a Flow Matching decoder for generation. (Inset: Spatial CNN architecture).}
    \label{fig2}
\end{figure*}

\noindent \textbf{Overview.}
As illustrated in Figure~\ref{fig2}, the EventSpeech framework orchestrates three synergistic modules to bridge the gap between asynchronous visual events and continuous acoustic waveforms. First, the Event Encoder transduces silent video into neuromorphic streams, extracting high-frequency kinematic features that encode precise articulatory dynamics. Second, the Multi-Scale Audio Encoder synergizes Selective State Space Models (SSM) with HWC. This hybrid architecture captures global prosodic context with linear complexity while preserving local spectral details, enabling the generation of rich timbre and intonation. The Hierarchical Alignment Module enforces synchronization at both the temporal (frame-level) and semantic (emotion-level) scales, ensuring the generated speech is not only lip-synced but also emotionally congruent with the visual input. Note that while visual events provide high-fidelity motion supervision during training, EventSpeech maintains multimodal consistency at inference, supporting both text-only generation via learned alignment priors and vision-augmented synthesis when events are available. 

\subsection{Neuromorphic Events Modeling}
\label{method1}
\noindent Given a silent video $V$ after audio--visual separation, where $V$ contains $T$ frames of size $H \times W \times 3$, we proposed \textit{Event Encoder} transduces static frames into a sparse events via a differentiable neuromorphic emulator V2E\cite{hu2021v2e}. Specifically, the simulator converts RGB inputs to grayscale and computes the log-luminance deviation between consecutive frames. Discrete events $E^{\pm}$ are triggered at coordinates $(x, y, t)$ precisely when this deviation surpasses learnable sensitivity thresholds $\{\theta^+, \theta^-\}$ for positive or negative polarities.

\noindent \textbf{Spatiotemporal Kinematic Encoding.}
To encode these sparse event signals into dense kinematic features, we employ a hierarchical architecture. A Spatial CNN\cite{woo2023convnext} first extracts local spatial textures, reducing the spatial dimension while increasing channel depth. Subsequently, bidirectional GRU (BiGRU) layers capture long-range temporal dependencies, yielding a spatiotemporal representation $H \in \mathbb{R}^{B \times T' \times 512}$, where $T' = T-1$ denotes the temporal resolution adjusted for interval-based kinematic modeling.

\noindent \textbf{Disentangled Visual Dynamics.}
To explicitly disentangle the multifaceted visual dynamics inherent in speech production, we introduce five specialized Multi-Head Feature Extractors (MHFE). These extractors project the shared high-frame-rate spatiotemporal features $H$ into distinct latent subspaces representing specific visual-speech attributes: \textit{Lip Motion}, \textit{Facial Action Units (AU)}, \textit{Head Pose}, \textit{Speaking Rhythm}, and \textit{Visual Prosody}. This extraction backbone is optimized via a hybrid supervision paradigm. Specifically, the explicit physical branches (Lip Motion, AU, and Head Pose) are guided by automated pseudo-labels derived from OpenFace. Conversely, for implicit temporal traits like Rhythm and Prosody that lack explicit labels, we enforce an orthogonality penalty alongside temporal modeling to achieve self-supervised disentanglement from the physical kinematics. To synthesize these heterogeneous cues, we employ a hierarchical fusion strategy. The aggregated temporal features are integrated with global speaker ($F_{spk}$) and emotion ($F_{emo}$) embeddings via a Multi-Layer Perceptron (MLP)\cite{tu2022maxim}. The resulting visual embedding $F_v$ synergizes microsecond-level kinematic precision with high-level semantic context, providing a robust condition. 

\subsection{Multi-Scale Audio Encoding }
\label{method2}
\noindent To extract high-fidelity acoustic representations from the reference audio $a$, we design the Audio Encoder, a distinct architecture that reconciles the trade-off between long-range temporal modeling\cite{gu2023mamba} and fine-grained spectral preservation. By synergizing selective SSM with the HWC, we efficiently encode global context with linear computational complexity while preserving fine-grained spectro-temporal nuances such as timbre and rhythm. Furthermore, we explicitly disentangle speech components through a dedicated prosody stream and a dual-path speech smoother, ensuring the final output $F_a$ is spectrally coherent and robust against generation artifacts.

\noindent \textbf{Acoustic Feature Decomposition.}
We first transform the raw waveform $a$ into a mel-spectrogram via Short-Time Fourier Transform (STFT)\cite{lee2022bigvgan}. Simultaneously, to explicitly encode prosodic variations, we extract frame-level pitch, energy, and duration contours. These components are individually quantized, embedded, and fused with the spectral features to construct the unified prosodic embedding $F_p \in \mathbb{R}^{B \times T_a \times D}$, serving as the composite input for subsequent modeling.

\noindent \textbf{Dual-Path Spectral Reﬁnement.}
To mitigate generation artifacts and enforce spectral smoothness, we introduce a dual-path mechanism. A 1D-convolutional branch extracts local temporal consistencies ($H_{\tau}$), while a parallel global linear branch preserves global spectral coherence ($H_{\omega}$). These complementary views are dynamically fused via a learnable gate $\alpha$. Subsequently, to incorporate speaker identity, a global timbre embedding $e_{\text{tmb}}$ is broadcasted along the temporal dimension and concatenated with the fused representation, resulting in the final acoustic representation $F_a$:
\begin{equation}
    F_a = W_f \left( \left[ \alpha \odot H_{\tau} + (1-\alpha) \odot H_{\omega} + \tilde{F} \right] \oplus e_{\text{tmb}} \right)
\end{equation}
where $\odot$ denotes element-wise multiplication, $\oplus$ represents channel-wise concatenation, and $W_f$ is a linear projection layer designed to integrate the multi-modal information. 

\noindent \textbf{Time-Frequency State Space Modeling.} 
Conventional Transformers suffer from quadratic complexity, limiting their efficacy on high-resolution audio. To address this, we employ an SSM-based architecture to efficiently model the global acoustic context of the prosodic embedding $F_p$. To complement this long-range temporal modeling and recover local spectral textures potentially attenuated by state-space representations, we integrate the proposed HWC for multi-scale time-frequency analysis. This synergistic design ensures the encoder captures global prosodic dependencies while retaining fine-grained formant structures, yielding the enhanced context features $\tilde{F}$.

\subsection{Hierarchical Cross-Modal Alignment }  
\label{method3}
\begin{figure}[htbp]
    \centering
    \includegraphics[width=0.65\columnwidth]{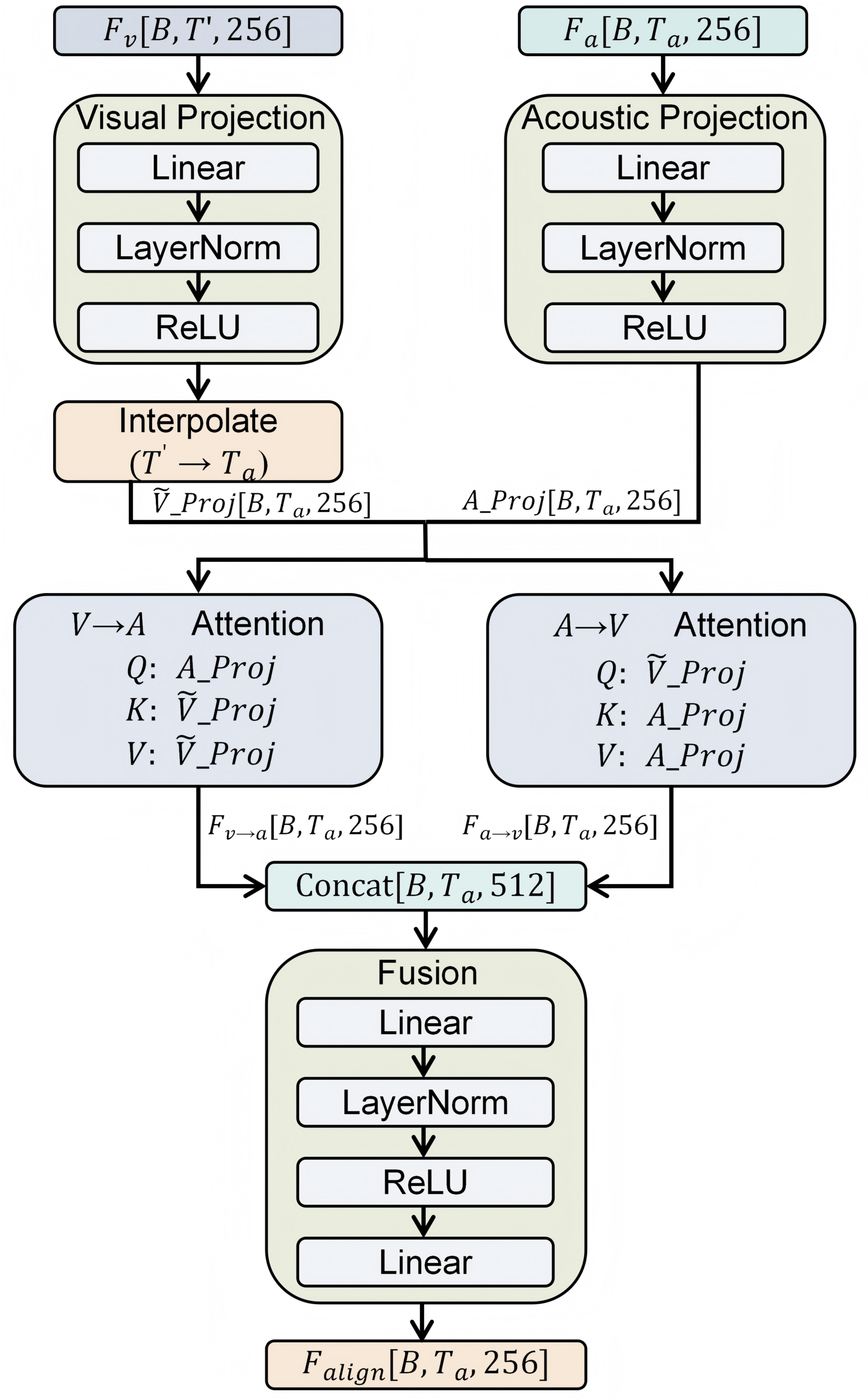}
    \caption{\textbf{Architecture of Hierarchical Cross-Modal Alignment}. Visual and acoustic features are projected and temporally aligned, then processed by bidirectional cross-attention (V$\rightarrow$A and A$\rightarrow$V). The concatenated outputs are fused via MLP to produce $F_{align}$.}
    \label{align}
\end{figure}

\noindent A critical challenge in event-driven speech generation is the inherent resolution mismatch between visual event frames ($T'$) and acoustic frames ($T_a$). We propose a Cross-Modal Alignment Module that synchronizes these modalities. The specific architectural design of this alignment mechanism is detailed separately in Figure \ref{align}. 

\noindent \textbf{Fine-Grained Temporal Registration.}
To bridge the inherent temporal granularity gap between visual event frames and acoustic frames, we first align their sequence lengths via interpolation\cite{copet2023simple}. Subsequently, we orchestrate a Bidirectional Cross-Attention mechanism to model dense, frame-level interactions. By allowing acoustic and visual representations to serve as reciprocal queries, this mechanism captures the intricate interplay between lip kinematics and phonetic content. The resulting bidirectional contexts are integrated via a linear projection, synthesizing a unified representation that ensures precise audio-visual synchronization.

\noindent \textbf{Emotion-Aware Manifold Alignment.}
Frame-level alignment alone may overlook high-level semantic consistency (e.g., emotional intensity). To enforce global coherence, we project the temporally pooled representations into a shared latent space, denoted $z_v = \Psi(\bar{F}_v)$ and $z_a = \Psi(\bar{F}_a)$. We then minimize the InfoNCE loss\cite{oord2018representation} to maximize mutual information between matched audio-visual pairs:
\begin{equation}
    \mathcal{L}_{\text{align}} = -\log \frac{\exp(\text{sim}(z_v, z_a) / \tau)}{\sum_{j=1}^{N} \exp(\text{sim}(z_v, z_a^j) / \tau)}
\end{equation}
where $z_a^j$ represents the acoustic embedding of the $j$-th negative sample, $\tau$ is the temperature parameter that scales the similarity distribution, and $N$ denotes the batch size. By jointly optimizing temporal and semantic alignment, our method generates speech that is not only lip-synced but also emotionally congruent, ensuring natural and expressive generation.

\subsection{Text Pipeline and Inference Architecture.}
\begin{figure}[H]
    \centering
    \includegraphics[width=0.75\columnwidth]{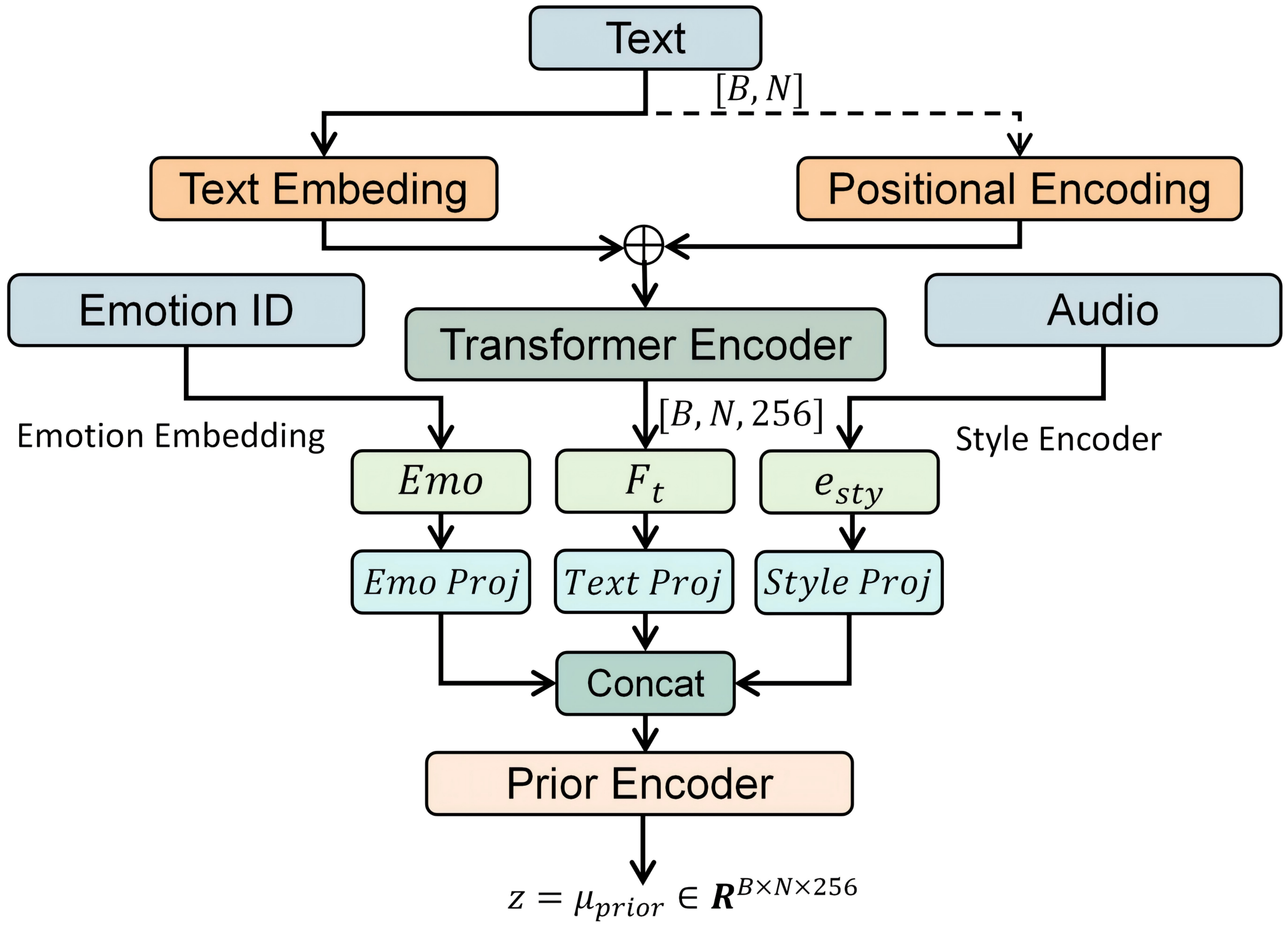}
    \caption{\textbf{Architecture of the Text Processing Backbone.} Text, emotion ID, and reference audio are encoded, projected and concatenated as the joint condition. The Prior Encoder predicts the distribution, with mean $\mu_{\text{prior}}$ serving as the latent representation $z$ for decoding. }
    \label{text1}
\end{figure}

\noindent Built upon the VITS\cite{kim2021conditional} dual stream paradigm EventSpeech establishes a rigorous mathematical connection between the training and inference latent spaces. During training a Posterior Encoder extracts the target latent representation $z_{post}$ directly from ground truth acoustic features. Concurrently the Prior Encoder processes the aligned multimodal conditions to predict the latent variable $z_{prior}$ explicitly minimizing the KL Loss to rigorously approximate the target posterior distribution.  During inference where ground truth audio is absent the framework entirely bypasses the posterior branch and exclusively relies on the trained Prior Encoder to generate the latent space. To seamlessly connect this dynamic architecture to the generation pipeline a dedicated Knowledge Bridge maps the active latent variable into unified hidden features. This mapping explicitly conditions an Optimal Transport Conditional Flow Matching (OT-CFM) decoder\cite{lipman2022flow} utilizing a Transformer-based architecture to solve the continuous time ordinary differential equation (ODE). This ultimately projects Gaussian noise onto the target mel spectrogram for waveform generation via a pretrained vocoder\cite{kong2020hifi}. Figure \ref{text1} supplements the main framework detailing for text processing. Specifically the joint condition concatenates the linguistic context $F_t$ the categorical emotion vector $Emo$ and the acoustic timbre $e_{sty}$ to explicitly guide the Prior Encoder. 

\section{BenchMark: EVT-SPK}
\label{EVT-SPK}

\noindent \textbf{Motivation.} Existing datasets primarily rely on RGB images or videos for speech generation\cite{kim2025faces,lee2023imaginary}.  While simulated events effectively mitigate data scarcity and resolve specific scenarios where standard RGB inputs suffer from severe motion blur, there remains a significant domain gap due to the lack of authentic physical noise and sensor-specific characteristics. To rigorously evaluate the capacity of native neuromorphic hardware in capturing richer kinematics constructing a physical dataset remains imperative. We therefore introduce the EVT-SPK benchmark which explicitly comprises a foundational synthetic corpus \textbf{EVT-SPK-Synth} and an authentic physical dataset \textbf{EVT-SPK-Real} to comprehensively validate our framework. 

\begin{figure}[H]
    \centering
    \includegraphics[width=\columnwidth]{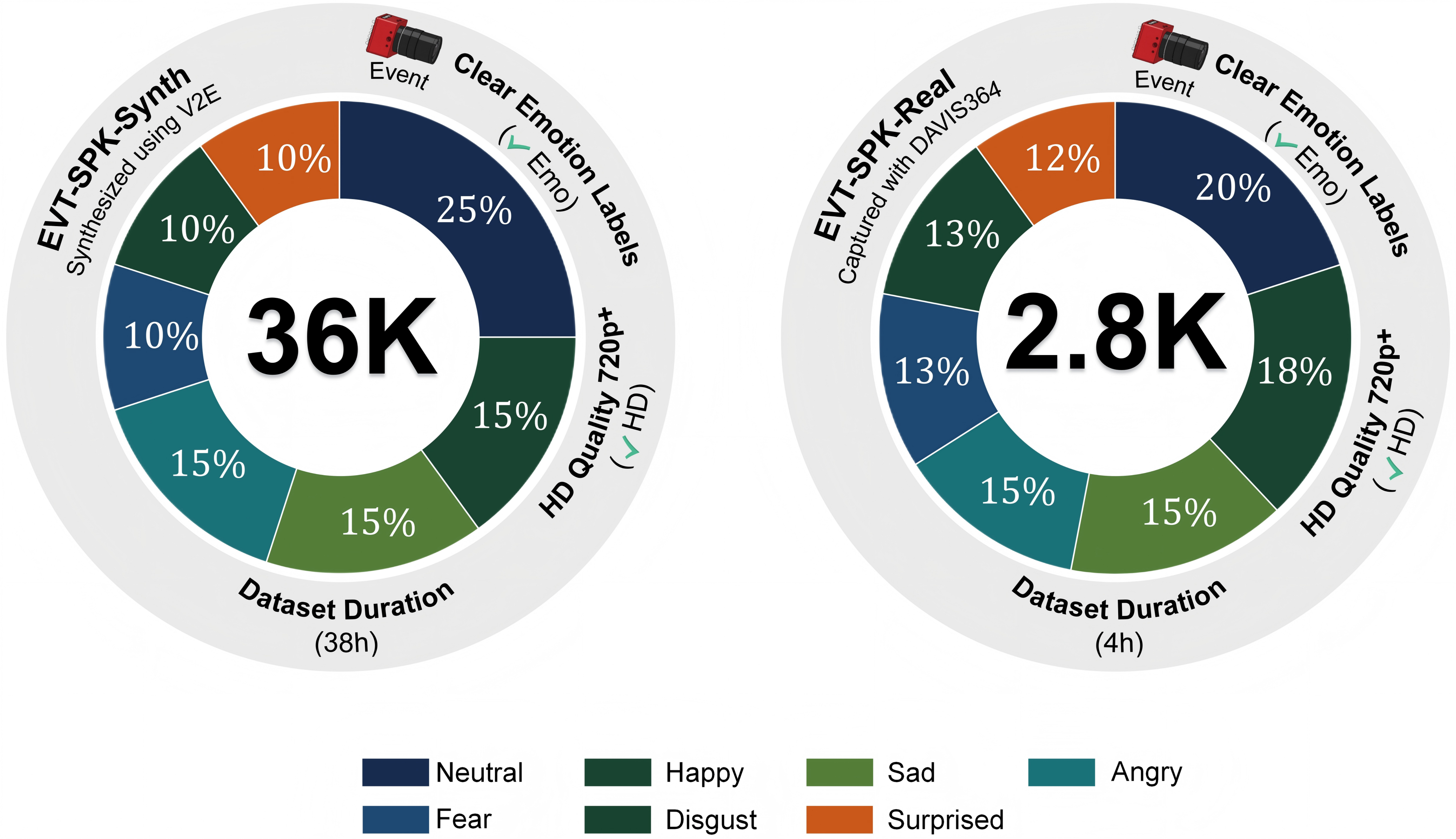}
    
    \caption{\textbf{Statistics of EVT-SPK.} EVT-SPK-Synth (\textbf{left}) comprises 36K clips (38 hours) synthesized via V2E. EVT-SPK-Real (\textbf{right}) contains 2.8K clips (4 hours) natively captured using a DAVIS346 camera. Both subsets guarantee HD visual quality alongside explicit emotion annotations.}
    \label{fig:dataset}
\end{figure}

\noindent \textbf{EVT-SPK-Synth.} We carefully select a high-quality audiovisual corpus rich in emotional expression based on a systematic selection criterion (Fig~\ref{fig:dataset}). We choose RAVDESS\cite{livingstone2018ryerson} and MEAD\cite{wang2020mead} because their respective advantages are valuable to our research: RAVDESS provides emotionally expressive speech performed by professional actors and follows a rigorous recording process; whilst MEAD provides more granular emotional intensity across a larger group of speakers. Although newer datasets like ESD\cite{zhou2021seen} and CMU-MOSEI\cite{zadeh2018multimodal} are larger in scale, they lack the synchronous high-quality video required for reliable event transitions, or primarily contain conversational speech rather than emotionally expressive speech. 

\begin{table*}[t]
\centering
\small
\caption{\textbf{Benchmarking results on the \textbf{EVT-SPK-Synth} and \textbf{EVT-SPK-Real} dataset sets between EventSpeech and baselines}. $^\dagger$: Audio ground truth is used for training. $^\S$: Converted into speech by adding timbre and emotion via AS. $^\Diamond$: The silent video is input as training. $^\ddagger$: Provided event signals through the V2E simulator. \textbf{T}: denotes the text-only variant without event stream conditioning. The best results achieved by our method are highlighted in {bold}.}
\label{tab2}
\renewcommand{\arraystretch}{1.15}
\resizebox{\textwidth}{!}{%
\begin{tabular}{@{} l *{9}{c} @{}}
\toprule
\textbf{Methods} & \textbf{MCD}{$\downarrow$} & \textbf{LSE-D}{$\downarrow$} & \textbf{LSE-C}{$\uparrow$} & \textbf{F0-RMSE}{$\downarrow$} & \textbf{MCD-SL}{$\downarrow$} & \textbf{KL}{$\downarrow$} & \textbf{WER}{$\downarrow$} & \textbf{CMOS}{$\uparrow$} & \textbf{SMOS}{$\uparrow$} \\
\midrule
\multicolumn{10}{@{}l@{}}{\textbf{\textit{EVT-SPK-Synth} (RAVDESS \& MEAD)}} \\
\quad Ground Truth (GT) & - & 7.12 & 0.856 & - & - & - & 0.024 & - & -  \\
\quad VALL-E 2$^\dagger$\cite{chen2024vall}  & 5.43 & 9.23 & 0.643 & 0.298 & 4.52 & 0.387 & 0.087 & $-1.12 \pm 0.21$ & $3.52 \pm 0.18$ \\
\quad MATCHA-TTS$^\dagger$\cite{mehta2024matcha}  & 4.82 & 8.76 & 0.698 & 0.234 & 3.98 & 0.298 & 0.062 & $-0.78 \pm 0.17$ & $3.68 \pm 0.15$ \\
\quad MMAudio+AS$^{\dagger\S}$\cite{cheng2025mmaudio}  & 5.89 & 9.87 & 0.587 & 0.342 & 4.87 & 0.456 & 0.108 & $-1.45 \pm 0.23$ & $3.15 \pm 0.19$ \\
\quad Diff-Foley+AS$^{\S\Diamond}$\cite{luo2023diff}  & 5.52 & 9.54 & 0.623 & 0.287 & 4.56 & 0.412 & 0.094 & $-1.28 \pm 0.20$ & $3.32 \pm 0.17$
\\
\quad VTS$^\Diamond$\cite{kim2025faces}  & 5.18 & 8.42 & 0.734 & 0.256 & 4.23 & 0.345 & 0.073 & $-0.92 \pm 0.18$ & $3.78 \pm 0.15$ 
\\
\quad VoiceCraft-Dub$^{\dagger\Diamond}$\cite{sung2025voicecraft}  & 4.35 & 8.15 & 0.762 & 0.215 & 3.85 & 0.254 & 0.058 & $-0.55 \pm 0.16$ & $3.98 \pm 0.14$ \\
\quad HPMDubbing$^{\dagger\Diamond}$\cite{cong2023learning}  & 6.25 & 8.95 & 0.655 & 0.355 & 5.12 & 0.485 & 0.148 & $-1.65 \pm 0.25$ & $3.25 \pm 0.21$ \\
\quad StyleDubber$^{\dagger\Diamond}$\cite{cong2024styledubber} & 5.65 & 8.52 & 0.715 & 0.315 & 4.65 & 0.415 & 0.112 & $-1.15 \pm 0.22$ & $3.55 \pm 0.18$ \\
\quad VTS+VE$^{\Diamond\ddagger}$\cite{kim2025faces}  & 4.76 & 8.18 & 0.756 & 0.223 & 3.89 & 0.276 & 0.067 & $-0.68 \pm 0.15$ & $3.92 \pm 0.14$ \\
\quad \textbf{EventSpeech-T (Ours)}& 3.89 & 8.08 & 0.744 & 0.179 & 3.12 &0.172 & 0.042 & $-0.52 \pm 0.13$ &$3.95 \pm 0.12$\\
\quad \textbf{EventSpeech (Ours)} & \textbf{3.67} & \textbf{7.56} & \textbf{0.843} &\textbf{ 0.156} & \textbf{2.94} & \textbf{0.126} & \textbf{0.038} & $\mathbf{-0.36 \pm 0.12}$ & $\mathbf{4.21 \pm 0.11}$ \\
\midrule
\multicolumn{10}{@{}l@{}}{\textbf{\textit{EVT-SPK-Real} (with DAVIS346)}} \\
\quad Ground Truth (GT) & - & 6.98 & 0.873 & - & - & - & 0.020 & - & -  \\
\quad VALL-E 2$^\dagger$\cite{chen2024vall}  & 5.08 & 8.74 & 0.662 & 0.271 & 4.21 & 0.358 & 0.081 & $-0.95 \pm 0.19$ & $3.75 \pm 0.16$ \\
\quad MATCHA-TTS$^\dagger$\cite{mehta2024matcha}  & 4.54 & 8.74 & 0.718 & 0.219 & 3.74 & 0.274 & 0.058 & $-0.55 \pm 0.16$ & $3.98 \pm 0.14$ \\
\quad MMAudio+AS$^{\dagger\Diamond}$\cite{cheng2025mmaudio}  & 5.51 & 9.28 & 0.608 & 0.318 & 4.54 & 0.421 & 0.101 & $-1.25 \pm 0.21$ & $3.35 \pm 0.18$ \\
\quad Diff-Foley+AS$^{\S\Diamond}$\cite{luo2023diff}  & 5.14 & 8.96 & 0.641 & 0.268 & 4.24 & 0.381 & 0.088 & $-0.88 \pm 0.19$ & $3.80 \pm 0.16$ \\
\quad VTS$^\Diamond$\cite{kim2025faces}  & 4.82 & 7.94 & 0.748 & 0.238 & 3.91 & 0.318 & 0.068 & $-0.75 \pm 0.17$ & $3.85 \pm 0.14$ \\
\quad VoiceCraft-Dub$^{\dagger\Diamond}$\cite{sung2025voicecraft}  & 3.95 & 7.65 & 0.795 & 0.185 & 3.45 & 0.215 & 0.045 & $-0.43 \pm 0.13$ & $4.18 \pm 0.12$ \\
\quad HPMDubbing$^{\dagger\Diamond}$\cite{cong2023learning}  & 5.85 & 8.55 & 0.685 & 0.325 & 4.85 & 0.445 & 0.125 & $-1.35 \pm 0.22$ & $3.35 \pm 0.19$ \\
\quad StyleDubber$^{\dagger\Diamond}$\cite{cong2024styledubber} & 5.25 & 8.15 & 0.745 & 0.285 & 4.35 & 0.365 & 0.088 & $-0.85 \pm 0.19$ & $3.72 \pm 0.16$ \\
\quad VTS+VE$^{\Diamond\ddagger}$\cite{kim2025faces}  & 4.38 & 7.68 & 0.771 & 0.207 & 3.54 & 0.254 & 0.062 & $-0.45 \pm 0.14$ & $4.10 \pm 0.13$ \\
\quad \textbf{EventSpeech-T (Ours)}& 3.56 & 7.81 & 0.764 & 0.165 & 3.07 &0.164 & 0.037 & $-0.44 \pm 0.12$ & $4.15 \pm 0.12$\\
\quad \textbf{EventSpeech (Ours)} & \textbf{3.18} & \textbf{7.28} & \textbf{0.843} & \textbf{0.124} & \textbf{2.41} &\textbf{0.108} & \textbf{0.028} & $\mathbf{-0.24 \pm 0.11}$  & $\mathbf{4.45 \pm 0.13}$ \\
\bottomrule
\end{tabular}%
}
\renewcommand{\arraystretch}{1.0}
\end{table*}
\noindent  \textbf{EVT-SPK-Real.} To validate framework efficacy we introduced this real-world physical subset capturing synchronized events use the \textit{DAVIS346} camera and premium audio through \textit{H3-VR} recorder maintaining strict synchronization $drift < 1ms$. Captured within a soundproof studio using DC driven LEDs to eliminate flicker this dataset comprises 4 hours of expressive speech from 15 actors across 7 emotions. Furthermore to explicitly demonstrate the superiority of events over standard RGB sensors we deliberately incorporated demanding physical scenarios encompassing extreme low illumination and rapid kinematic dynamics inducing severe motion blur.   

\section{Experiment and Results}
\label{experiment}

\subsection{Setup}

\noindent \textbf{Implementation Details.} We train the 113M parameter EventSpeech on most of the EVT-SPK-Synth dataset using 6 NVIDIA A100 GPUs. Optimization is performed via AdamW with OneCycleLR\cite{kim2021conditional} scheduling for approximately 940K iterations. To ensure robustness against real-world variations, we apply multi-modal augmentations, including temporal jittering for events and standard spectral perturbations for audio. At inference, we utilize a 20-step flow matching solver\cite{lipman2022flow} coupled with a fine-tuned HiFi-GAN vocoder\cite{kong2020hifi}, generating 8-second audio clips with a Real-Time Factor (RTF) of 0.006 (approx. 48ms per clip). 

\noindent \textbf{Baselines.} As the first framework to explore neuromorphic event-driven speech generation, we establish a comprehensive set of competitive baselines across three related domains. 
\textbf{(i) Text-to-Speech: }(1) VALL-E 2\cite{chen2024vall}, a SOTA neural codec language model achieving human-like zero-shot TTS, and (2) MATCHA-TTS\cite{mehta2024matcha}, utilizing conditional flow matching for efficient generation. 
\textbf{(ii) Video-to-Audio:} Since standard video-to-audio baselines generate environmental sounds rather than intelligible speech, we introduce an Audio-to-Speech (AS). Using self-supervised representations\cite{qian2022contentvec}, it extracts disentangled linguistic content to translate raw acoustic output into clear speech waveforms. This yields (3) MMAudio+AS\cite{cheng2025mmaudio} and (4) Diff-Foley+AS\cite{luo2023diff}, adapting SOTA latent diffusion sound models for our specific task. 
\textbf{(iii) Visual Voice Cloning \& Dubbing:} To represent recent advances in visually-conditioned speech, we evaluate (5) VTS\cite{kim2025faces}, utilizing hierarchical visual representations; (6) VoiceCraft-Dub\cite{sung2025voicecraft}, a neural codec model for strict lip-synced generation; (7) HPMDubbing\cite{cong2023learning}, which extracts hierarchical facial and scene prosody; and (8) StyleDubber\cite{cong2024styledubber}, emphasizing multi-scale phoneme-level style learning. 
Furthermore, to provide a direct cross-modal comparison, we evaluate (9) VTS+VE\cite{kim2025faces}, which equips the VTS architecture with a Video-to-Events (VE) simulator designed to synthesize events from standard RGB video.

\begin{figure*}[t]
    \centering
    \includegraphics[width=\linewidth]{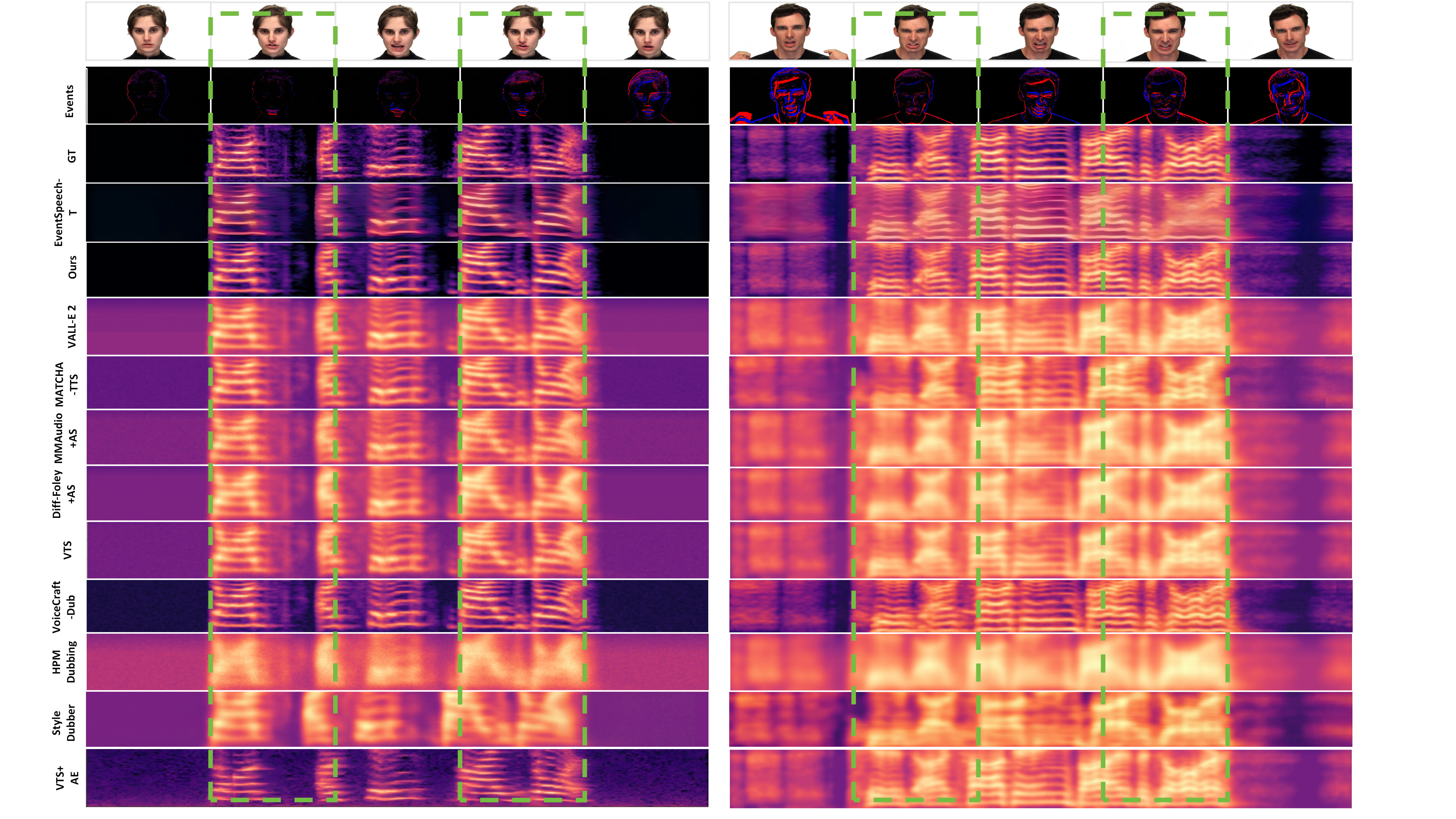}
    \caption{\textbf{Qualitative comparison of generated Mel-spectrograms between our method and competitive baselines}. The highlight regions where EventSpeech exhibits superior capability in synthesizing fine-grained emotional fluctuations, natural prosodic transitions, and temporal consistency aligned with the GT.}
    \label{fig:3}
\end{figure*}

\subsection{Evaluation Metrics}
\noindent We conduct the comprehensive speech evaluation using multiple dimensions to compare our method and baselines. \\
\noindent \textbf{Acoustic Fidelity.} DTW-aligned Mel-Cepstral Distortion (MCD) quantifies spectral distortion against ground-truth references while mitigating temporal mismatches \cite{li2023styletts}.\\
\noindent \textbf{Temporal Synchronization.} SyncNet metrics \cite{niu2023audio,prajwal2020lip,yaman2024audio} including Lip Sync Error-Distance (LSE-D) and Confidence (LSE-C) measure audio-visual alignment assisted by ground-truth streams to verify strict onset synchronization.\\
\noindent \textbf{Prosodic Naturalness.} F0-RMSE on log-transformed fundamental frequency trajectories \cite{jiang2024mega,sankar2024indicvoices} alongside Speaker-Level normalized MCD (MCD-SL) \cite{li2025styletts} jointly assess prosody-aware acoustic quality.\\
\noindent \textbf{Emotional Preservation.} To quantify emotional fidelity, we compute the Kullback-Leibler (KL) Divergence\cite{copet2023simple} between the emotional-prosodic feature distributions.\\
\noindent\textbf{Speech Intelligibility.} We evaluate Word Error Rate (WER) using the Whisper-large-v3 ASR model\cite{kim2024paralinguistics} to assess linguistic content accuracy from inference input text.\\
\noindent \textbf{Subjective Metrics.} We conduct human evaluations following standard protocols\cite{chen2024vall,ju2024naturalspeech} utilize Comparative MOS (CMOS) and Speaker Similarity MOS (SMOS) to respectively assess perceptual quality and identity preservation.

\subsection{Main Results}
\noindent The benchmarking results across both the EVT-SPK-Synth and EVT-SPK-Real datasets validate the advanced capabilities and robust generalization of the proposed EventSpeech framework. To ensure a comprehensive evaluation, we assess all models utilizing their official inference implementations under standard protocols. Our comparative analysis encompasses three distinct categories of baselines including pure TTS systems such as VALL-E 2\cite{chen2024vall} and MATCHA-TTS\cite{mehta2024matcha}, specialized video dubbing frameworks like VoiceCraft-Dub\cite{sung2025voicecraft} and StyleDubber\cite{cong2024styledubber}, and conventional RGB-based video-to-speech approaches including VTS\cite{kim2025faces}, MMAudio+AS\cite{cheng2025mmaudio}, Diff-Foley+AS\cite{luo2023diff}, and VTS+VE\cite{kim2025faces}. Furthermore, we explicitly evaluate a purely text-driven variant denoted as EventSpeech-T across both datasets. This text-only baseline is introduced to independently validate our generative architecture and isolate the performance gains directly attributable to neuromorphic events. As detailed in Table \ref{tab2}, EventSpeech consistently achieves SOTA performance across all evaluated acoustic and alignment metrics under diverse conditions. 

\noindent\textbf{EVT-SPK-Synth Evaluation.} On the synthetic dataset, EventSpeech demonstrates EventSpeech's superior performance on the synthetic dataset, surpassing SOTA methods across all dimensions. The text-only EventSpeech-T maintains highly competitive acoustic fidelity and prosodic naturalness to effectively exceed several vision-conditioned models. The subsequent integration of neuromorphic events into the complete architecture further elevates performance in metrics sensitive to fine-grained dynamics. This empirical advantage indicates that event representations explicitly isolate high-frequency articulatory dynamics. Consequently, our formulation facilitates precise cross-modal alignment even within simulated environments by effectively disentangling essential motions from the redundant spatial textures inherent in standard frames. 

\noindent\textbf{EVT-SPK-Real Evaluation.} EventSpeech demonstrates robust performance on the challenging real-world dataset. The purely text-driven EventSpeech-T experiences an expected degradation in lip synchronization due to the complete absence of visual physiological constraints. While conventional baselines like VTS+VE\cite{kim2025faces} achieve competitive alignment scores likely because standard RGB representations remain unaffected by realistic sensor degradation, the complete EventSpeech framework successfully leverages physical events to overcome these limitations. By capturing richer physiological dynamics directly from the physical sensor, the complete architecture significantly outperforms both its text-only counterpart and all external competitors. The full model secures the highest overall LSE-C score to demonstrate highly precise cross-modal synchronization while achieving an unparalleled MCD of 3.18 and an F0-RMSE of 0.124. 

\subsection{Ablations}
\noindent For the ablation study, we use the EVT-SPK-Synth dataset, and each study is executed based on the 113M model. The default configuration of our full model is highlighted in \textbf{Bold}. 
 \begin{table}[H]
    \centering
\renewcommand{\arraystretch}{1.1}
   \small
   \footnotesize
    \caption{\textbf{Impact of high-speed cameras against event}.}
    \label{albation1}
    \begin{tabular}  {lcccc}\toprule
       \textbf{Variants} & \textbf{MCD$\downarrow$} &\textbf{ LSE-C$\uparrow$} & \textbf{F0-RMSE$\downarrow$} & \textbf{KL$\downarrow$} \\
        \midrule
        RGB 25 FPS & 5.43 & 0.685 & 0.248 & 0.231 \\
        RGB 60 FPS     & 4.58 & 0.762 & 0.195 & 0.184 \\
        RGB 120 FPS   & 4.12 & 0.794 & 0.173 & 0.159 \\
        \textbf{Ours} & \textbf{3.67} & \textbf{0.843} & \textbf{0.156} & \textbf{0.126} \\ 
        \bottomrule
    \end{tabular}
\end{table}

\noindent \textbf{Events versus High-Speed Cameras.} Table \ref{albation1} demonstrates that performance gains plateau at 120 fps since conventional cameras remain bound by the \textit{exposure-readout} paradigm and inevitably suffer from intra-frame motion blur during rapid articulation. Conversely, EventSpeech achieves optimal metrics including an MCD of 3.67 and an LSE-C of 0.843 by transcending discrete sampling. This confirms the fundamental advantage of neuromorphic vision in capturing continuous biological dynamics with microsecond precision to preserve fast articulatory cues lost even in high-speed videos.

\noindent \textbf{Efficacy of HWC Architecture.} We validate the synergy between the Mamba and Wavelet modules within the HWC backbone. As reported in Table \ref{albation2}, the variant excluding both modules struggles with timbre fidelity. Mamba captures global prosodic dependencies via selective SSM while Wavelets preserve local spectral hierarchies. Their combination yields the lowest MCD to ensure robust timbre reconstruction.

\begin{table}[htbp]
    \centering
    \renewcommand{\arraystretch}{1.1}
    \small
    \footnotesize
    \caption{\textbf{Impact of Hierarchical Wavelet Contextualizer}.}
    \label{albation2}
    \begin{tabular}{lcccc}\toprule
        
        % 第一行添加灰色底色，\rowcolor参数：gray!30 表示30%灰度（可调整数值）
         \textbf{Variant}& \textbf{MCD$\downarrow$}&\textbf{ LSE-C$\uparrow$}&\textbf{ F0-RMSE$\downarrow$}& KL$\downarrow$\\\midrule
        
        w/o Both& 5.90& 0.653& 0.316& 0.368\\
        Mm& 4.34& 0.755& 0.207& 0.209\\
        Wet& 4.71& 0.725& 0.231& 0.241\\
      \textbf{Mm+Wet}& \textbf{3.67}& \textbf{0.843}& \textbf{0.156}& \textbf{0.126}\\ \bottomrule
    \end{tabular}
\end{table}
\noindent \textbf{Cross-Modal Alignment.} As reported in Table \ref{albation3}, naive concatenation and linear interpolation (LinInterp) fail to model non-linear coarticulation, causing severe temporal misalignment. Similarly, unidirectional attention (UndiAtten) suffers from inherent visual ambiguities. Conversely, our Bidirectional Cross-Attention (Bi-DiAtten) establishes reciprocal reasoning whereby visual features temporally anchor acoustic boundaries and acoustic context resolves visual ambiguities. This mutual constraint cultivates a tightly coupled latent manifold, evidenced by minimized KL divergence and superior LSE-C scores, to ensure highly synchronous speech generation.

\begin{table}[htbp]
    \centering
    \renewcommand{\arraystretch}{1.1}
   \small
   \footnotesize
   
    \caption{\textbf{Impact of Cross-Modal Alignment with vision and speech}.}
    \label{albation3}
    \begin{tabular}{lcccc}\toprule
        
        % 第一行添加灰色底色，\rowcolor参数：gray!30 表示30%灰度（可调整数值）
        \textbf{Variant}& \textbf{MCD$\downarrow$}&\textbf{ LSE-C$\uparrow$}&\textbf{ F0-RMSE$\downarrow$}& \textbf{KL}$\downarrow$\\\midrule
        
        w/o Align& 5.62& 0.623& 0.301& 0.321\\
        LinInterp& 4.55& 0.711& 0.226& 0.239\\
        UndiAtten& 4.07& 0.770& 0.176& 0.164\\
      \textbf{Bi-DiAtten}& \textbf{3.67}& \textbf{0.843}& \textbf{0.156}& \textbf{0.126}\\ \bottomrule
    \end{tabular}
\end{table}

\noindent \textbf{Voxelization versus Native Events.} Table \ref{tab:voxel_bins} indicates that coarse voxelization ($N=1$) degrades acoustic fidelity by discarding microsecond-level sparsity, whereas processing dense events ($N=6$) induces severe computational bottlenecks in GFLOPs and RTF. Our optimal binning ($N=3$) resolves this dilemma by leveraging mature deep learning operations to preserve asynchronous temporal advantages. This pragmatic compromise achieves a superior balance between computational efficiency and synthesis quality, yielding competitive MCD and LSE-C scores with drastically reduced overhead.
\begin{table}[htbp]
    \centering
    \caption{\textbf{Impact of voxelization on native events}.}
    \label{tab:voxel_bins}
    \footnotesize 
    \renewcommand{\arraystretch}{1.1} 
    \setlength{\tabcolsep}{3pt} 
    \begin{tabular}{@{}lcccccc@{}}
        \toprule
      \textbf{Voxel Bins} &\textbf{ MCD$\downarrow$} & \textbf{LSE-C$\uparrow$} &\textbf{ F0-RMSE$\downarrow$} &\textbf{ KL$\downarrow$} &\textbf{ GFLOPs$\downarrow$} & \textbf{RTF$\downarrow$} \\
        \midrule
        N=1 (Coarse)        & 3.92 & 0.781 & 0.158 & 0.142 & 8.6  & 0.12 \\
        \textbf{N=3 (Ours)} & \textbf{3.18} &\textbf{ 0.843} & \textbf{0.124} & \textbf{0.108} & \textbf{11.2} & \textbf{0.16} \\
        N=6 (Native)        & 3.06 & 0.851 & 0.119 & 0.105 & 16.4 & 0.23 \\
        \bottomrule
    \end{tabular}
\end{table}

\section{Conclusion}
\label{conclusion}
\noindent In this work, we propose \textbf{EventSpeech} to address the inherent Temporal Granularity Mismatch in conventional frame-based speech synthesis at the physical level by introducing neuromorphic events. We design the MHFE and HWC architectures to precisely capture microsecond-level facial kinematics while optimizing global prosody and local timbre, effectively overcoming the limitations of existing approaches. Furthermore, we construct \textbf{EVT-SPK} as the first comprehensive benchmark comprising both simulated events and natively captured real-world data. Extensive evaluations demonstrate that our method achieves state-of-the-art performance against competitive baselines. We anticipate this work will establish a robust new paradigm for expressive neuromorphic speech generation.

\section{Discussion}
\noindent While EventSpeech pioneers neuromorphic speech synthesis, certain inherent limitations warrant further rigorous investigation. The limited EVT-SPK-Real scale forces reliance on simulated events missing the complex dynamics of physical sensor noise. Furthermore, practical voxelization compromises raw temporal sparsity, and insufficient extreme condition data restricts our proposed framework's universal generalization. Addressing these constraints, our future trajectory focuses on scaling the physical dataset and investigating Neural Ordinary Differential Equations to enable continuous modeling, ultimately bridging the simulation-to-reality gap.

%\newpage
%\vfill
\nocite{*}
\bibliographystyle{ieeetr}
\bibliography{references}

%\end{IEEEbiographynonphoto}
\end{document}